\documentclass{iaus}
\usepackage{graphicx}

\title[Age \& metallicity of Galactic clusters] 
{Age and metallicity of Galactic clusters from full spectrum fitting}

\author[Koleva et al.]   
{Mina Koleva$^{1,2}$%
  \thanks{mina.koleva@obs.univ-lyon1.fr},
 Philippe Prugniel$^{1,3}$, Pierre Ocvirk$^4$, Damien Le~Borgne$^4$}

\affiliation{
$^1$Universit\'e Lyon~1,
Observatoire de Lyon, St. Genis
Laval, 69230, France; CNRS UMR 5574;\\[\affilskip]
$^2$Department of Astronomy, St. Kl. Ohridski University of Sofia, BG-1164 Sofia, Bulgaria;\\[\affilskip]
$^3$Observatoire de Paris, GEPI, F-75014, France;\\[\affilskip]
$^4$CEA Saclay/Service d'Astrophysique, Gif-sur-Yvette Cedex, F-91191, France;\\[\affilskip]
}

\pubyear{2007}
\volume{241}  
\pagerange{1-2}
\date{30 Jan 2007}
\jname{Proceedings Stellar Populations as Building Blocks of Galaxies}
\editors{A. Vazdekis \& R. Peletier, eds.}
\begin{document}

\maketitle

\begin{abstract}
We are using full spectrum fitting to determine the ages and metallicities
of Galactic clusters (M67 and globular clusters). 
We find that the method is very accurate to measure the metallicity. 
Blue horizontal branches are
well identified as a 'young' sub-component, and the age of the 'old'
component is in fair agreement with CMD determinations.

\end{abstract}

\noindent {\bf Goal.}
The spectroscopic determination of age and metallicity of star clusters
is a key to understand the characteristics and origin of
extragalactic cluster systems. It is also a challenge for analysis
methods. The goal of this work is to analyze spectra of Galactic clusters
for which CMD measurements are independently available.

\vskip 2pt
\noindent {\bf Observational material.}
We are using the spectrum of the open cluster M67 synthesized
by Schiavon et al (2004) and the library of spectra of globular
clusters assembled by Schiavon et al. (2005). The spectra have
a resolution of about 3~$\AA $ and the analysis is performed in
the optical range between H\&K and H$_\alpha$.

\vskip 2pt
\noindent {\bf Method.}
Since our goal is to study extragalactic clusters and dwarf galaxies,
we are concerned by the optimal usage of the information. For this
reason we use full spectrum fitting rather than spectroscopic indices.
Another reason is that modern observations provide relatively high
resolution spectra of clusters (R=1000-5000) that would not be adequately 
exploited using indices. 
Still, as well as indices, our method is designed to be essentially insensitive
to the shape of the continuum, so that the characteristics of the population
are constrained by the relative depths and shapes of the spectral lines.
The method consists in comparing observed spectra to population models
generated with Pegase.HR (Le~Borgne et al. 2004) using the library 
ELODIE.3.1 (in preparation, see Prugniel \& Soubiran 2001, 2004 for the
previous versions).
We are using two programs implementing this method: STECKMAP 
(Ocvirk et al. 2006), can constrain the history of the stellar population, and
NBursts (Chilingarian, this conference) determines the 
SSP-equivalent parameters. Both programs measure simultaneously
the broadening of the spectrum (kinematical and observational) and
the characteristics of the population.
The precision of the method has been discussed in Koleva et al. (2006)
and consistency tests are presented in Koleva et al. (this conference).

\begin{figure}
\centering
\resizebox{6.5cm}{!}{\includegraphics{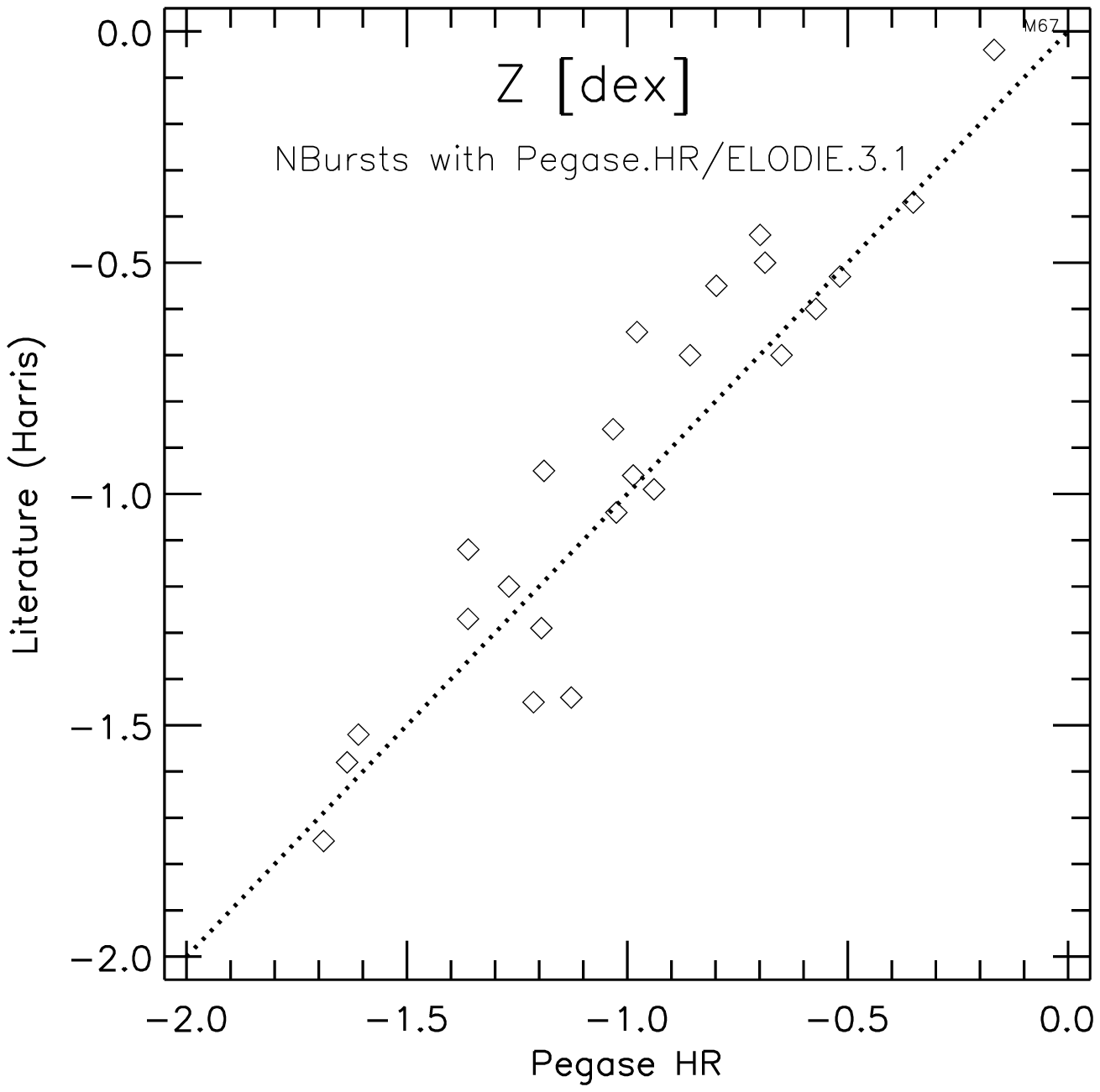} }
\resizebox{6.5cm}{!}{\includegraphics{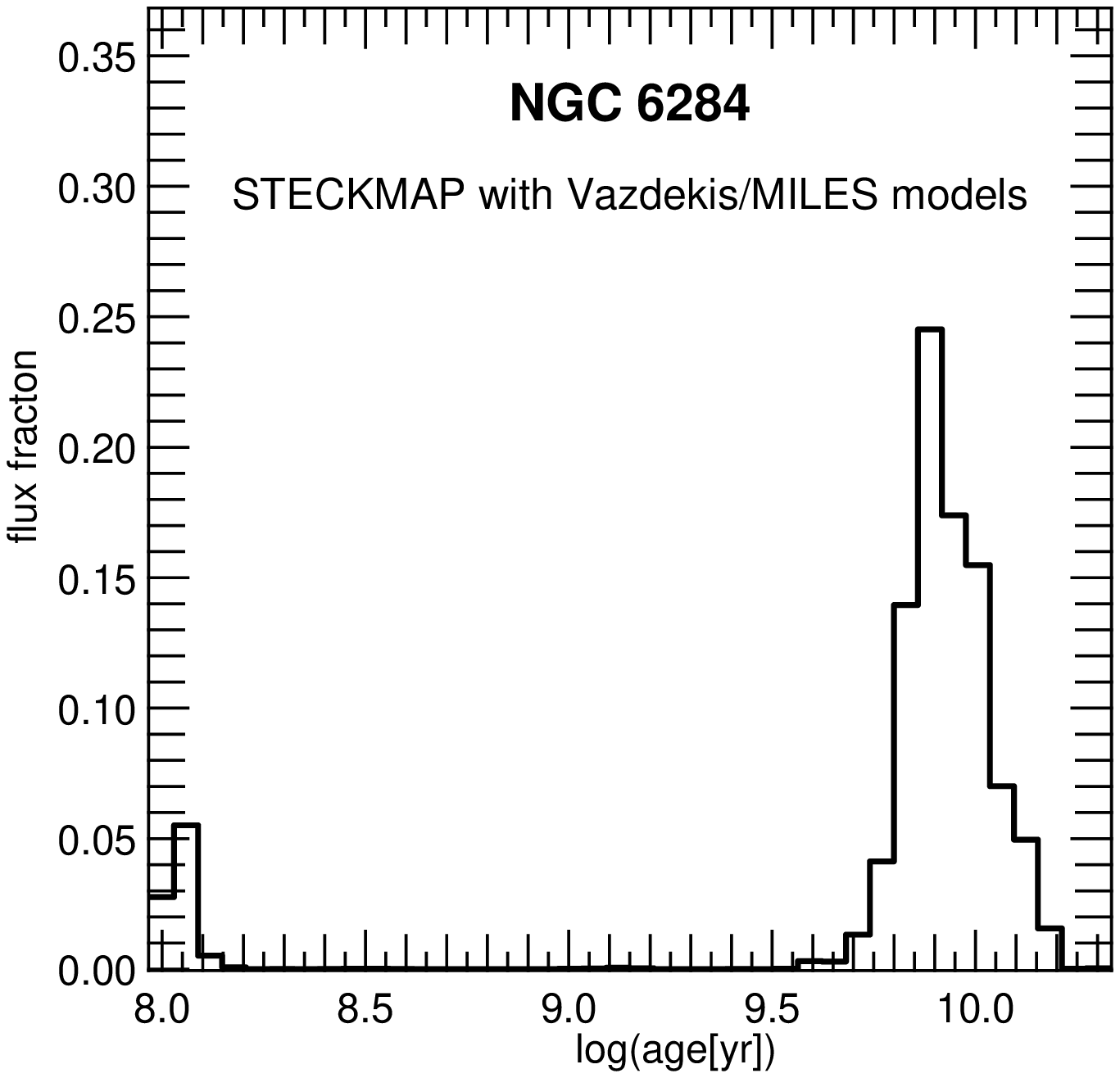} }
\caption {(a) Comparison between metallicity measured by spectrum fitting and 
literature. (b) Detection of 'young' component in BHB clusters: case of NGC~6284.}
\label{Fig1}
\end{figure}

\vskip 2pt
\noindent {\bf Results.}
The quality of the fits is in general excellent, and
does not appear to be limited by the model, except for high metallicity
clusters where $\alpha$-element abundances exceed the models (see Prugniel et al.,
this conference). The SSP-equivalent metallicities (Fig.~1a) are in good
agreement with CMD determination and have a precision is 0.15 dex.

The SSP-equivalent ages are strongly affected by the HB
morphology: Clusters with BHB are found younger with
NBursts.
This excess of blue stars is
detected as a 'young subcomponent' in the stellar age distributions
reconstructed with STECKMAP (Fig.~1b). 
The age of the 'old' component agrees better with CMD fits
than the SSP-equivalent age.

The ages of the old stellar systems are very sensitive 
to the temperature scale of the cold giants. A difference of 200K changes the
age estimate by 20 or 30\%.

\vskip 2pt
\noindent {\bf Conclusion.}
Spectrum fitting of globular clusters robustly pinpoint BHBs (or blue stragglers), and it seems possible to fit a model with a BHB component.

Full spectrum fitting, with our method,
is a reliable approach to find metallicities and ages of
stellar systems over a wide range of parameters. Using the high spectral 
resolution models built with
ELODIE.3.1 or with CFLIB (Valdes et al. 2004) will allow us to understand
 better the origin of extragalactic globular cluster systems.

\vskip 2pt
\noindent {\bf Acknowledgements}
We thanks R. Schiavon for giving the M67 spectrum. MC acknowledges the financial support from IAU. DLB acknowledges the French CNES.

\end{document}